\documentclass[preprint, superscriptaddress,pra,aps,showpacs]{revtex4-1}

\usepackage{graphicx}
\usepackage{amsmath,amsfonts,
}
\usepackage{amssymb}
\usepackage{bm}% bold math
\begin{document}

\title{Nonlinear Acoustics FDTD method  \\ including Frequency Power Law Attenuation \\ for Soft Tissue Modeling}

\author{No\'e Jim\'enez}
\affiliation{Instituto de Investigaci\'on para la Gesti\'on Integrada de Zonas Costeras, Universitat Polit\`ecnica de Val\`encia, Paranimf 1, 46730 Grao de Gandia, Spain}

\author{Javier Redondo}
\affiliation{Instituto de Investigaci\'on para la Gesti\'on Integrada de Zonas Costeras, Universitat Polit\`ecnica de Val\`encia, Paranimf 1, 46730 Grao de Gandia, Spain}

\author{V\'ictor S\'anchez-Morcillo}
\affiliation{Instituto de Investigaci\'on para la Gesti\'on Integrada de Zonas Costeras, Universitat Polit\`ecnica de Val\`encia, Paranimf 1, 46730 Grao de Gandia, Spain}

\author{Francisco Camarena}
\affiliation{Instituto de Investigaci\'on para la Gesti\'on Integrada de Zonas Costeras, Universitat Polit\`ecnica de Val\`encia, Paranimf 1, 46730 Grao de Gandia, Spain}

\author{Yi Hou }
\affiliation{Department of Biomedical Engineering, Columbia University, New York, NY USA}

\author{Elisa E. Konofagou}
\affiliation{Department of Biomedical Engineering, Columbia University, New York, NY USA}
\affiliation{Department of Radiology, Columbia University, New York, NY USA}

\begin{abstract} 
This paper describes a model for nonlinear acoustic wave propagation through absorbing and weakly dispersive media, and its numerical solution by means of finite differences in time domain method (FDTD). The attenuation is based on multiple relaxation processes, and provides frequency dependent absorption and dispersion without using computational expensive convolutional operators. In this way, by using an optimization algorithm the coefficients for the relaxation processes can be obtained in order to fit a frequency power law that agrees the experimentally measured attenuation data for heterogeneous media over the typical frequency range for ultrasound medical applications. Our results show that two relaxation processes are enough to fit attenuation data for most soft tissues in this frequency range including the fundamental and the first ten harmonics. Furthermore, this model can fit experimental attenuation data that do not follow exactly a frequency power law over the frequency range of interest. The main advantage of the proposed method is that only one auxiliary field per relaxation process is needed, which implies less computational resources compared with time-domain fractional derivatives solvers based on convolutional operators.
\end{abstract}


\maketitle

\section{Introduction}
Accurate prediction of finite amplitude acoustic waves traveling through biological media are essential in developing new therapy and imaging techniques for medical ultrasound applications. Numerous experimental studies show that attenuation $\alpha (f)$ of biological media exhibits a power law dependence on frequency $\alpha(f)=\alpha _0 f^{\gamma}$ over the frequency range used in medical applications \cite{Hill2004} \cite{Duck2012}. In this sense, considering an initial monochromatic plane wave traveling through nonlinear media, two opposite effects govern the final wave amplitude: on one hand, higher harmonics appear and their amplitude grows as a consequence of nonlinear progressive wave steepening; and on the other hand, the damping for each harmonic is different, following the above power law. Therefore, differences in the frequency power law model can lead to huge differences in attenuation of higher harmonics, and hence the inclusion of frequency dependence attenuation is critical for correctly predict nonlinear propagation.

Classical thermo-viscous \cite{Pierce1989} sound attenuation exhibit a squared frequency dependence, $\gamma=2$, however the empirically fitted power law for soft tissues typically ranges between $\gamma=(0.6,2) $\cite{Hill2004} \cite{Duck2012} \cite{Goss1979}. For many tissues the attenuation shows a dependence that can be modeled by $\gamma$ close to the unity, where the value varies from different tissues. Moreover, for some individual examples the local value of $\gamma$ exhibit lower values at low ultrasound frequencies and tends to 2 in the high frequency limit \cite{Hill2004}. Despite the study of the physical mechanism besides this complex frequency dependence is out of the scope of this work, there exist numerous phenomenological approaches for including the observed losses in the acoustic equations\cite{Wismer1995} \cite{Kellya2009}. On the other hand, is common in literature to describe the losses in soft tissues as multiple-relaxation processes \cite{Pierce1989} \cite{Nachman1990} \cite{Hill2004}, where the relaxation frequencies can be associated to either a tissue specific physical mechanism or empirically optimized to fit the observed tissue attenuation \cite{Cleveland1995}, \cite{Pinton2009}. Moreover, fractional partial differential operators has been demonstrated the ability to describe frequency power law attenuation \cite{Szabo1994} \cite{Prieur2011}. These operators can be included in the modeling by means of time \cite{Szabo1994}, space \cite{Chen2004} or combined time-space fractional derivatives \cite{Caputo1967} \cite{Wismer2006}. The latter operators can be biological motivated and derived from tissue micro-structure using fractal ladders based on networks of springs and dashpots \cite{Kellya2009}. However, these fractional loss operators can be also derived from a continuum of relaxation process \cite{Nasholm2011}, that suggest that the two approaches can be equivalent under certain conditions \cite{Treeby2012}.

In order to solve these models, many time-domain numerical methods have been developed. Attenuation modeled by relaxation processes can be solved by means of finite-differences in time-domain (FDTD) solvers in linear regime \cite{Yuan1999} and in nonlinear regime applied to augmented Burger's equation \cite{Cleveland1995}, Khokhlov-Zabolotskaya-Kuznetsov (KZK) \cite{Yang2005} and Westervelt \cite{Pinton2009} nonlinear wave equations. On the other hand, time-dependent fractional derivatives can be solved in nonlinear regime \cite{Liebler2004} by convolutional operators. This approach requires the memory storage of certain time history, and although the memory can be strongly reduced compared to direct convolutions, this algorithm employs up to ten auxiliary fields and a memory buffer of three time steps. In order to overcome this limitation, time-space fractional derivatives or fractional Laplacian \cite{Treeby2010} \cite{Chen2004} can be used to model frequency power laws without time-domain convolutional operators. Recently k-space and pseudo-spectral methods have been applied in order to solve fractional Laplacian operators efficiently in nonlinear regime \cite{Treeby2012}.

The aim of this work is to present a numerical method that solves the complete set of equations (continuity, momentum and state equations) including nonlinear propagation and frequency power law attenuation based in multiple relaxation processes. The numerical method presented here is based on the FDTD method. The inclusion of relaxation processes in the presented formulation avoids convolutional operators so only one extra field per relaxation process is needed and no memory buffer is needed. The paper is organized as follows: in Sec. \ref{s:model} the model equations that describes the problem are exposed, Sec. \ref{s:numerical} describes the computational method presented in this work and in Sec. \ref{s:validation} the method is validated comparing the numerical results with known analytic solutions for linear and nonlinear regimes.

\section{Physical model}\label{s:model}
\subsection{Full-wave modeling}\label{s:model:nonlinear}
The principles of mass and momentum conservation lead to the main constitutive relations for nonlinear acoustic waves, which for a fluid can be expressed as \cite{Naugolnykh1998}
\begin{equation}\label{eq:continuity}
	\frac{{\partial \rho }}{{\partial t}} =  - \nabla  \cdot \left( {\rho {\bf{v}}} \right)
\end{equation}
and
\begin{equation}\label{eq:momentum}
\rho \left( {\frac{{\partial {\bf{v}}}}{{\partial t}} + {\bf{v}} \cdot \nabla {\bf{v}}} \right) =  - \nabla p + \eta {\nabla ^2}{\bf{v}} + \left( {\zeta  + \frac{\eta }{3}} \right)\nabla \left( {\nabla  \cdot {\bf{v}}} \right),
\end{equation}
where $\rho$ is the total density field,$\textbf{v}$ is the particle velocity vector,$p$ is the pressure,$\eta$ and $\zeta$ are the coefficients of shear and the bulk viscosity respectively. The acoustic waves described by this model exhibit viscous losses with squared power law dependence on frequency. In order to include a power law frequency dependence on the attenuation, a multiple relaxation model will be added into the time domain equations.

The basic mechanism for energy loss in a relaxing media is the appearance of a phase shift between the pressure and density fields. This behavior is commonly modeled as a time dependent connection at the fluid state equation, that for a fluid retaining the nonlinear effects up to second order an be expressed as \cite{Naugolnykh1998}\cite{Rudenko1977}:
\begin{equation}\label{eq:st_cont}
	p = c_0^2\rho ' + \frac{{c_0^2}}{{{\rho _0}}}\frac{B}{{2A}}{\rho '^2} + \int_{ - \infty }^t G (t - t')\frac{{\partial \rho '}}{{\partial t}}dt,
\end{equation}
where $\rho '=\rho - \rho _0$ is the density perturbation over the stationary density $\rho _0$, $B/A$ is the nonlinear parameter, $c_0$ is the small amplitude sound speed, and $G(t)$ is the kernel associated with the relaxation mechanism. The first two terms describe the instantaneous response of the medium and the convolutional third term accounts for the ``memory time" of the relaxing media. Thus, by choosing an adequate time function for the kernel $G(t)$ the model can present an attenuation and dispersion response that fits the experimental data of the heterogeneous media. However, the direct resolution of the constitutive relations (\ref{eq:continuity}-\ref{eq:st_cont}) in this integral form is a complex numerical task due to the convolutional operator. Thus, instead of describe $G(t)$ with a specific time domain waveform, the response of the heterogeneous medium can be alternatively described by a sum of $N$ relaxation processes with exponential time dependence as:
\begin{equation}\label{eq:relax_define}
	\int_{ - \infty }^t G (t - t')\frac{{\partial \rho '}}{{\partial t}}dt = \sum\limits_{n = 1}^N {{G_n}} * \frac{{\partial \rho '}}{{\partial t}},
\end{equation}
with the $n$-th order kernel expressed as
\begin{equation}\label{eq:define_G}
	{G_n}(t) = {\eta _n}c_0^2{e^{\frac{{ - t}}{{{\tau _n}}}}}H(t),
\end{equation}
where $H(t)$ is the Heaviside piecewise function $H(t < 0) = 0$, $H(t > 0) = 1$, $\tau _n$ is the characteristic relaxation time and $\eta _n$ the relaxation parameter for the $n$-th order process. This last dimensionless parameter controls the amount of attenuation and dispersion for each process as $\eta _n = (c_n^2 - c_0^2)/c_0^2$, where $c_n$ is the sound speed in the high frequency limit associated to $n$-th order relaxation process, also known as the speed of sound in the ``frozen" state \cite{Pierce1989}. In order to describe relaxation without the need of including a convolutional operator, we shall define a state variable $S_n$ for each process as
\begin{equation}\label{eq:Sn}
	{S_n} = \frac{1}{\tau _n}{G_n} * \rho '
\end{equation}
Thus, using the convolutional property $\frac{\partial }{{\partial t}}\left( {G(t) * \rho '(t)} \right) = \frac{{\partial G(t)}}{{\partial t}} * \rho '(t) = G(t)*\frac{{\partial \rho '(t)}}{{\partial t}}$, the time derivative of the relaxation state variable obeys the following relation for the n-th order process:
\begin{equation}\label{eq:dSndt_long}
	\frac{{\partial {S_n}}}{{\partial t}} = \left( { - \frac{1}{{{\tau _n}}}\frac{{{\eta _n}c_0^2}}{{{\tau _n}}}{e^{ - \frac{t}{{{\tau _n}}}}}H(t) + \frac{{{\eta _n}c_0^2}}{{{\tau _n}}}{e^{ - \frac{t}{{{\tau _n}}}}}\delta (t)} \right) * \rho '
\end{equation}
where $\delta (t)$ is the Dirac delta function. Using the Eq. (\ref{eq:Sn}) this relation becomes a simple ordinary differential equation for each process as:
\begin{equation}\label{eq:dSdt}
	\frac{{\partial {S_n}}}{{\partial t}} =  - \frac{1}{{{\tau _n}}}{S_n} + \frac{{{\eta _n}c_0^2}}{{{\tau _n}}}\rho '
\end{equation}
Using again convolutional properties, we can substitute Eq. (\ref{eq:dSdt}) into (\ref{eq:relax_define}), and the relaxing nonlinear state Eq. (\ref{eq:st_cont}) becomes:
\begin{equation}\label{eq:st_final_c0}
	p = c_0^2\rho ' + \frac{{c_0^2}}{{{\rho _0}}}\frac{B}{{2A}}{\rho '^2} - \sum\limits_{n = 1}^N {{S_n}}  + \sum\limits_{n = 1}^N {{\eta _n}c_0^2\rho '} 
\end{equation}
Moreover, if ``frozen" sound speed for $N$ mechanisms is defined as $c_\infty ^2 = c_0^2\left( {1 + \sum\limits_{n = 1}^N {{\eta _n}} } \right)$, Eq. (\ref{eq:st_final_c0}) leads to:
\begin{equation}\label{eq:st_final_cinf}
	p = c_\infty ^2\rho ' + \frac{{c_0^2}}{{{\rho _0}}}\frac{B}{{2A}}{\rho '^2} - \sum\limits_{n = 1}^N {{S_n}}.
\end{equation}
Due to the smallness of the relaxation parameter, $\eta _n$, i.e. when weak dispersion is modeled, the sound speed in the high frequency limit reduces to \cite{Naugolnykh1998}:
\begin{equation}\label{eq:cinf}
	{c_\infty } = {c_0}\left( {1 + \sum\limits_{n = 1}^N {\frac{{{\eta _n}}}{2}} } \right)
\end{equation}
Note Eq. (\ref{eq:st_final_cinf}) for a mono-relaxing media is equivalent to that can be found in literature \cite{Rudenko1977}
\begin{equation}\label{eq:st_cinf}
	p = c_\infty ^2\rho ' + \frac{{c_0^2}}{{{\rho _0}}}\frac{B}{{2A}}{\rho '^2} - \int\limits_{ - \infty }^t {\frac{{\eta c_0^2}}{\tau }} {e^{ - \left( {\frac{{t - t'}}{\tau }} \right)}}\rho '(t').
\end{equation}
Thus, the constitutive relations to solve by means of the numerical method in the nonlinear regime are the continuity Eq. (\ref{eq:continuity}), the motion Eq. (\ref{eq:momentum}) and the second order fluid state relaxing Eq. (\ref{eq:st_final_cinf}), where the state variable $S_n$ obeys the relation (\ref{eq:dSdt}) for the $n$-th order relaxation process. Although the aim of this work is to model biological media, the generalized formulation presented here can be used to describe the attenuation and hence the dispersion observed in other relaxing media, as the relaxation processes of oxygen and nitrogen molecules in air or the relaxation associated with boric acid and magnesium sulfate in seawater \cite{Pierce1989}.

\subsection{Small amplitude modeling}\label{s:model:linear}
On the other hand, if small amplitude perturbations are considered, an equivalent derivation of this model can be expressed for multiple relaxation media \cite{Yuan1999}. Thus, for an homogeneous inviscid relaxing fluid the linearized continuity and motion Eq. (\ref{eq:continuity}-\ref{eq:momentum}) reduces to
\begin{equation}\label{eq:lin_cont}
	\frac{{\partial \rho }}{{\partial t}} =  - {\rho _0}\nabla  \cdot {\bf{v}}
\end{equation}
and
\begin{equation}\label{eq:lin_momen}
	{\rho _0}\frac{{\partial {\bf{v}}}}{{\partial t}} =  - \nabla p;
\end{equation}
and linearizing the fluid state Eq. (\ref{eq:st_final_cinf}) we obtain:
\begin{equation}\label{eq:lin_st}
	\rho ' = \frac{1}{{c_\infty ^2}}\left( {p + \sum\limits_{n = 1}^N {{S_n}} } \right)
\end{equation}
These equations can be solved directly in this form, however, if expressed in pressure-velocity formulation the density field is no longer necessary and computational effort can be reduced. Thereby, assuming a linear ``instantaneous" compressibility $\kappa _\infty = {\rho _0}c_\infty ^2$, and substituting Eq. (\ref{eq:lin_st}) into Eq. (\ref{eq:lin_cont}) yields
\begin{equation}\label{eq:lin_p1}
	\frac{{\partial p}}{{\partial t}} + \sum\limits_{n = 1}^N {\frac{{\partial {S_n}}}{{\partial t}}}  =  - {\kappa _\infty }\nabla  \cdot {\bf{v}}
\end{equation}
Then, taking the time derivative of the state variable Eq. (\ref{eq:dSdt}) we get
\begin{equation}\label{eq:lin_p2}
	\frac{{\partial p}}{{\partial t}} - \sum\limits_{n = 1}^N {\frac{1}{{{\tau _n}}}} {S_n} + \rho '\sum\limits_{n = 1}^N {\frac{{{\eta _n}c_0^2}}{{{\tau _n}}}}  =  - {\kappa _\infty }\nabla  \cdot {\bf{v}}
\end{equation}
Finally, substituting again the linearized state Eq. (\ref{eq:lin_st}) and arranging terms the linearized continuity equation leads to
\begin{equation}\label{eq:lin_pconti}
	\frac{{\partial p}}{{\partial t}} + p\sum\limits_{n = 1}^N {\frac{{{\eta _n}c_0^2}}{{{\tau _n}c_\infty ^2}}}  + \sum\limits_{n = 1}^N {\left( {\frac{{{\eta _n}c_0^2}}{{{\tau _n}c_\infty ^2}} - \frac{1}{{{\tau _n}}}} \right){S_n}}  =  - {\kappa _\infty }\nabla  \cdot {\bf{v}}.
\end{equation}
On the other hand, the state evolution equation can be expressed as a function of the acoustic pressure as
\begin{equation}\label{eq:lin_dSndt}
	\frac{{\partial {S_n}}}{{\partial t}} =  - \frac{1}{{{\tau _n}}}{S_n} + \frac{{{\eta _n}c_0^2}}{{{\tau _n}c_\infty ^2}}\left( {p - \sum\limits_{n = 1}^N {{S_n}} } \right).
\end{equation}
Thus, the linearized governing Eq. (\ref{eq:lin_momen},\ref{eq:lin_pconti}) for a relaxing media are expressed in a pressure-velocity formulation and can be solved together with the coupled state evolution equation (\ref{eq:lin_dSndt}) by means of standard finite differences numerical techniques \cite{Yuan1999}. In this way, lossless linear acoustics equations can be obtained by setting $\eta _n =0$ or in the limit when the relaxation times ${\tau _n} \to \infty $. The relaxation behavior described by this linearized model is achieved too by the formulation described in Ref. \cite{Yuan1999}, where the relaxation coefficients $\eta _n$ and the relaxation variable $S_n$ are defined in a different, but analogous way.

\section{Computational method}\label{s:numerical}
\subsection{Discretization}\label{s:numerical:discr}
Equations (\ref{eq:continuity}, \ref{eq:momentum}, \ref{eq:st_final_cinf}, \ref{eq:dSdt}) are solved by means of finite-differences in time-domain (FDTD) method. Cylindrical axisymmetric ${\bf{x}} = (r,z)$ coordinate system is considered in this work, however, the method can be derived in other coordinate system. As in the standard acoustic FDTD method \cite{Botteldooren1996}, the particle velocity fields are discretized staggered in time and space respect to the density and pressure fields, Fig. 1. Uniform grid is considered, where $r=i\Delta r$, $z=j\Delta z$, $t=m \Delta t$, with $\Delta r$ and $\Delta z$ as the radial and axial spatial steps, and $\Delta t$ is the temporal step.
\begin{figure}[htbp]
	\centering
	\includegraphics[width=8cm]{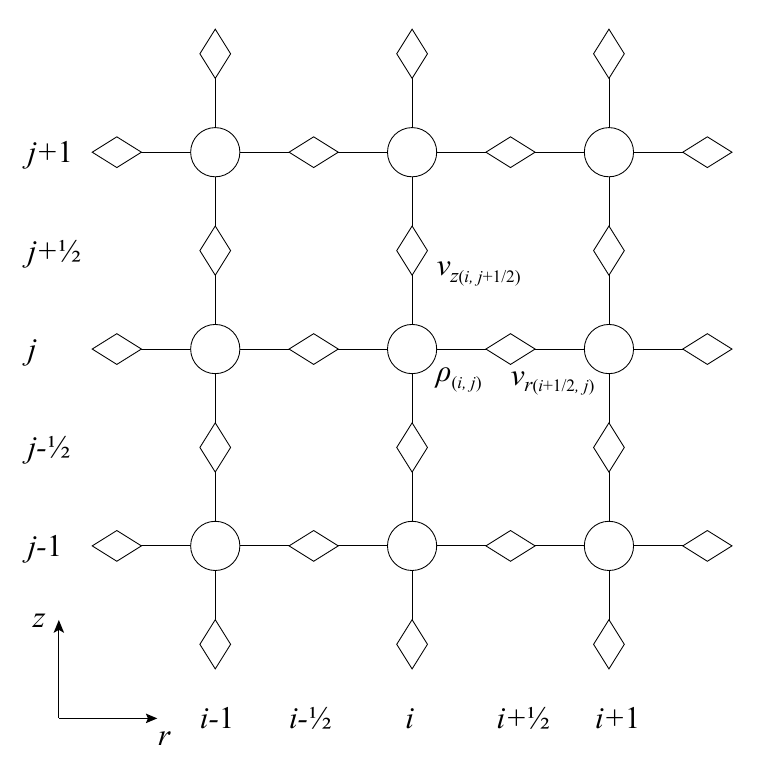}
	\caption{Spatial discretization as in standard acoustics FDTD method. The pressure $(p_{i,j}^m)$ and the $n$-th order relaxation process state fields $(S_{n,i,j}^m)$ are evaluated at same discrete location as the density $(\rho _{i,j}^m)$. Particle velocity fields are discretized staggered in both space and time respect to the density, pressure and the $n$-th order state fields.}
	\label{fig1_fdtd_grid}
\end{figure}
Thus, centered finite differences operators are applied over the partial derivatives of the governing equations. In order to fulfill the conservation principles over each discrete cell of the domain interpolation is needed over the off-center grid variables \cite{LeVeque1992}. The $r$ component of Eq. (\ref{eq:momentum}) is expressed in a cylindrical axisymmetric system as 
\begin{align}\label{eq:mot_r}
	\frac{{\partial {v_r}}}{{\partial t}} = &- \frac{1}{\rho}\frac{{\partial p}}{{\partial r}} -  {{v_r}\frac{{\partial {v_r}}}{{\partial r}} - {v_z}\frac{{\partial {v_r}}}{{\partial z}}} \\
 \nonumber & + \frac{\eta}{\rho} \left( {\frac{{{\partial ^2}{v_r}}}{{\partial {r^2}}} + \frac{1}{r}\frac{{\partial {v_r}}}{{\partial r}} + \frac{{{\partial ^2}{v_r}}}{{\partial {z^2}}} - \frac{{{v_r}}}{{{r^2}}}} \right) \\
 \nonumber & + \frac{1}{\rho}\left( { \zeta + \frac{1}{3}\eta } \right)\left( {\frac{{{\partial ^2}{v_r}}}{{\partial {r^2}}} + \frac{1}{r}\frac{{\partial {v_r}}}{{\partial r}} + \frac{{{\partial ^2}{v_z}}}{{\partial r\partial z}} - \frac{{{v_r}}}{{{r^2}}}} \right).
\end{align}
In the present method, each term of the above expression is approximated by centered finite differences evaluated at $r = (i+{1\over 2})\cdot\Delta r$, $z = (j+{1\over 2})\cdot\Delta z$, $t = (m+\tfrac{1}{2})\cdot\Delta t$. This equation can be solved obtaining an update equation for ${v_r}_{i+{1\over 2}, j}^{m+{1\over 2}}$. In the same way, the $z$ component of the motion equation (\ref{eq:momentum}) is expressed as
\begin{align}\label{eq:mot_z}
	\frac{{\partial {v_z}}}{{\partial t}} =  &  - \frac{1}{\rho}\frac{{\partial p}}{{\partial z}} -  {{v_r}\frac{{\partial {v_z}}}{{\partial r}} - {v_z}\frac{{\partial {v_z}}}{{\partial z}}} \\
  \nonumber &+ \frac{\eta}{\rho} \left( {\frac{{{\partial ^2}{v_z}}}{{\partial {r^2}}} + \frac{1}{r}\frac{{\partial {v_z}}}{{\partial r}} + \frac{{{\partial ^2}{v_z}}}{{\partial {z^2}}}} \right)\\
  \nonumber &+ \frac{1}{\rho}\left( {\zeta  + \frac{1}{3}\eta } \right)\left( {\frac{{{\partial ^2}{v_r}}}{{\partial z\partial r}} + \frac{1}{r}\frac{{\partial {v_r}}}{{\partial z}} + \frac{{{\partial ^2}{v_z}}}{{\partial {z^2}}}} \right)
\end{align}
Each term of this expression is approximated by centered finite differences and evaluated at $r=i\cdot\Delta r$, $z=(j+{1\over 2})\cdot \Delta z$, $t= m\cdot\Delta t$. An update equation is obtained solving this equation for ${v_z}_{i,j+{1\over 2}}^{m + {1\over 2}}$. The equation (\ref{eq:continuity}) in cylindrical axisymmetric coordinate system is expressed as
\begin{equation}\label{eq:cont_cyl}
	\frac{{\partial \rho }}{{\partial t}} = -\rho \left( {\frac{{\partial {v_r}}}{{\partial r}} + \frac{{{v_r}}}{r} + \frac{{\partial {v_z}}}{{\partial z}}} \right) - {v_r}\frac{{\partial \rho }}{{\partial r}} - {v_z}\frac{{\partial \rho }}{{\partial z}}.
\end{equation}
Following the same procedure, each term of the above expression is approximated by centered finite differences and evaluated at $r = i\cdot\Delta r$, $z = j\cdot\Delta z$, $t = (m+{1\over 2})\cdot\Delta t$, and the update equation is obtained solving this expression for $\rho _{i,j}^{m + 1}$. Finally, the equation (\ref{eq:dSdt}) is solved for $m+1$ as
\begin{equation}\label{eq:st_discrete}
	S_{n,i,j}^{m + 1} = S_{n,i,j}^{m} + \Delta t\left( -\frac{1}{\tau _n}S_{n,i,j}^{m} +\frac{\eta _n c_0^2}{\tau _n}\rho '^{m}_{n,i,j} \right).
\end{equation}

\subsection{Solver}\label{s:numerical:solver}
In order to fulfill the conservation principles for all the discrete cells an averaging of the off centered values is proposed \cite{LeVeque1992}, and the temporal averaging of the fields leads to an implicit system of equations. However, to obtain an explicit solution, the equations can be solved using an iterative solver. In this way, the future field values depend not only in the past values but also on unknown future values:
\begin{equation}\label{eq:solver1}
	{v_r}_{i + \frac{1}{2},j}^{m + \frac{1}{2}} = f\left( {{v_r}_{i + \frac{1}{2},j}^{m'},{v_r}_{i + \frac{1}{2},j}^{m - \frac{1}{2}},{v_z}_{i,j + \frac{1}{2}}^{m - \frac{1}{2}},\rho _{i,j}^m,p_{i,j}^m} \right)
\end{equation}
\begin{equation}\label{eq:solver2}
	{v_z}_{i,j + \frac{1}{2}}^{m + \frac{1}{2}} = f\left( {{v_z}_{i,j + \frac{1}{2}}^{m'},{v_r}_{i + \frac{1}{2},j}^{m - \frac{1}{2}},{v_z}_{i,j + \frac{1}{2}}^{m - \frac{1}{2}},\rho _{i,j}^m,p_{i,j}^m} \right)
\end{equation}
\begin{equation}\label{eq:solver3}
	\rho _{i,j}^{m + 1} = f\left( {\rho _{i,j}^{m' + \frac{1}{2}},\rho _{i,j}^m,{v_r}_{i + \frac{1}{2},j}^{m - \frac{1}{2}},{v_z}_{i,j + \frac{1}{2}}^{m - \frac{1}{2}}} \right)
\end{equation}
The fields in the above expressions with temporal index $m'$ are the temporal averaged values and are calculated by the iterative solver as:
\begin{equation}\label{eq:iterate1}
	{v_r}_{i + \frac{1}{2},j}^{m'} = \frac{{{v_r}_{i + \frac{1}{2},j}^{m - \frac{1}{2}} + {v_r}_{i + \frac{1}{2},j}^{m + \frac{1}{2}}}}{2}
\end{equation}
\begin{equation}\label{eq:iterate2}
	{v_z}_{i,j + \frac{1}{2}}^{m'} = \frac{{{v_z}_{i,j + \frac{1}{2}}^{m - \frac{1}{2}} + {v_z}_{i,j + \frac{1}{2}}^{m + \frac{1}{2}}}}{2}
\end{equation}
\begin{equation}\label{eq:iterate3}
	\rho _{i,j}^{m' + \frac{1}{2}} = \frac{{\rho _{i,j}^m + \rho _{i,j}^{m + 1}}}{2}
\end{equation}
The initial value for the iterative solver is obtained by means of the explicit FDTD linear solution. This value is injected into the implicit system (\ref{eq:solver1}-\ref{eq:solver3}). Due to the fact that the linear value is very similar to the nonlinear solution for a small time step, the iterative algorithm converges fast. In all the simulations presented in the validation section one iteration is used. Finally, a leap-frog time marching is applied to solve each time step until the desired simulation time is reached.

\subsection{Boundary conditions}\label{s:numerical:bc}
Perfectly matched layers (PML) \cite{Liu1997} were placed in the limits of the domain ($\pm z$ and $+r$) to avoid spurious reflections from the limits of the integration domain. Inside the PML domains linearized acoustics equations were solved using the complex coordinate screeching method \cite{Liu1999}. These absorbent boundary conditions have reported an attenuation coefficient of 55.2 dB for a layer of 30 elements and broadband wave with 1 MHz central frequency and non-normal incidence angle. The staggered grid is terminated on velocity nodes, so the coupling to the PML layers is implemented by only connecting these external nodes. This allows us to prevent the singularity of the cylindrical coordinate system: due to the staggered grid, the only variable discretized at $r=0$ is $v_r$, and axisymmetric condition ${v_r}{|_{r = 0}} = 0$ is applied there. To solve spatial differential operators on boundaries some ``phantom" nodes must be created with the conditions:
\[\begin{array}{*{20}{l}}
{{v_r}( - r)}&{ =  - {v_r}(r),}\\
{{v_z}( - r)}&{ = {v_z}(r),}\\
{\rho ( - r)}&{ = \rho (r),}\\
{p( - r)}&{ = p(r).}
\end{array}\]

\subsection{Stability}\label{s:numerical:stablity}
The stability for the lossless algorithm follows the Courant-Friedrich-Levy (CFL) condition, so for uniform grid $(\Delta r = \Delta z = \Delta h)$ the maximum duration of the time step is limited by $\Delta t \le \Delta h/{c_0}\sqrt D $ where $D$ is the number of dimensions (in our case, cylindrical axisymmetric coordinate system, $D=2$). However, numerical instabilities have been observed when ${\tau _f}/2\pi < \Delta t$. Due to this empirical relation, the maximum values for relaxation frequencies are limited too by the chosen spatial discretization by the simple relation
\begin{equation}
	{f_n} < \sqrt 2 {N_\lambda }{f_0},
\end{equation}
where ${f_n} = 2\pi /{\tau _n}$ is the maximum relaxation frequency for all processes, ${N_\lambda }$ is the number of elements per wavelength and $f_0$ the frequency of the propagating wave. At higher amplitudes nonlinear effects induce the growing of harmonics of the fundamental frequency of the initial wave. The diffusive viscosity term in the motion equation, Eq. (\ref{eq:momentum}), attenuates the small wavenumbers, damping the ``node to node" numerical oscillations and ensuring numerical stability in weakly nonlinear regime. However, when discontinuities are present in the solution, extra numerical techniques must be employed to guarantee convergence. In this work artificial viscosity \cite{Ginter2002} was added when shock waves are present in the solution. Thus, by choosing the artificial viscosity as $\eta \propto \Delta {h^2}$ the algorithm shows consistency when $\Delta h \to 0$, so if stability is achieved by the CFL condition, the convergence is guaranteed. Thereby, the viscosity term ensures the entropy condition is fulfilled \cite{Ginter2002}, and the method is able to solve shock waves in viscous fluids.

\section{Validation}\label{s:validation}
\subsection{Single relaxation process}\label{s:validation:relax}
In order to validate the frequency dependent attenuation and dispersion of the numerical method a simulation was done in linear regime including a single relaxation process. A homogeneous medium was considered, where the parameters were set to the typical values for water at 20 ºC: $c_0=1500$ m/s, $\rho _0=1000$ kg/m$^3$, $B/A=5$, $\eta=8.90 \cdot 10^{-4}$ Pa$\cdot$s. A single relaxation process was included, with a characteristic relaxation time of ${\tau _1} = 1/2\pi {f_0}$ and relaxation modulus of $\eta _1=0.0134$ that leads to a frozen sound speed of $c_\infty=1510$ m/s. In this case, the numerical parameters were set to $\Delta r=\Delta z=1.87\cdot 10^{-7}$ m and $\Delta t=8.65 \cdot 10^{-11}$ s. A plane wave front traveling in $+z$ direction was considered. Thus, at $z=0$ the media was excited with a negative second derivative of a Gaussian function:
\begin{equation}\label{eq:ricker}
	p(t,{z_0}) = {p_0}\left( {1 - 2{{\left( {\pi {f_0}t} \right)}^2}} \right){e^{ - {{\left( {\pi {f_0}t} \right)}^2}}}.
\end{equation}
Here, the central frequency of the broadband signal was set to $f_0=1$ MHz, and the amplitude was set to $p_0=1$ $\mu$ Pa, small enough to neglect the nonlinear propagation terms. As the wave propagates, a single relaxation process change the complex amplitude as $\exp{(-az)}$, where $a$ can be expressed as \cite{Naugolnykh1998}
\begin{equation}\label{eq:acomplex}
	a = \frac{{\omega {\eta _1}(i{\omega ^2}\tau _1^2 - \omega {\tau _1})}}{{2{c_0}(1 + {\omega ^2}\tau _1^2)}}.
\end{equation}
Thus, the theoretical attenuation for the relaxation processes was estimated from the real part of $a$, and including viscous losses leads to
\begin{equation}\label{eq:alpha_t}
	\alpha (\omega ) = \frac{{{\omega ^2}}}{{2{\rho _0}c_0^3}}\left( {\zeta  + \frac{4}{3}\eta } \right) + \sum\limits_{n = 1}^N {\frac{{{\eta _n}}}{{2{c_0}{\tau _n}}}\frac{{{\omega ^2}\tau _n^2}}{{1 + {\omega ^2}\tau _n^2}}} .
\end{equation}
In order to compute the attenuation of the numerical method, simulated pressure was recorded at two locations $z_0$ and $z_1$, and attenuation and phase velocity were estimated from the spectral components over the bandwidth of the input signal. The numerical attenuation was calculated as
\begin{equation}\label{eq:alpha_num}
	\alpha {(\omega)_{fdtd}} = \frac{{\ln \left( {\left| {P(\omega,{z_1})/P(\omega,{z_0})} \right|} \right)}}{{\left( {{z_1} - {z_0}} \right)}}
\end{equation}
where $P$ is the Fourier transform of the measured pressure waveforms at points $z_0$ and $z_1$. The results, plotted in Fig. 2, show excellent agreement between the attenuation obtained by numerical solution and the predicted by the theory for a single relaxation process.

\begin{figure}[htbp]
	\centering
	\includegraphics[width=8cm]{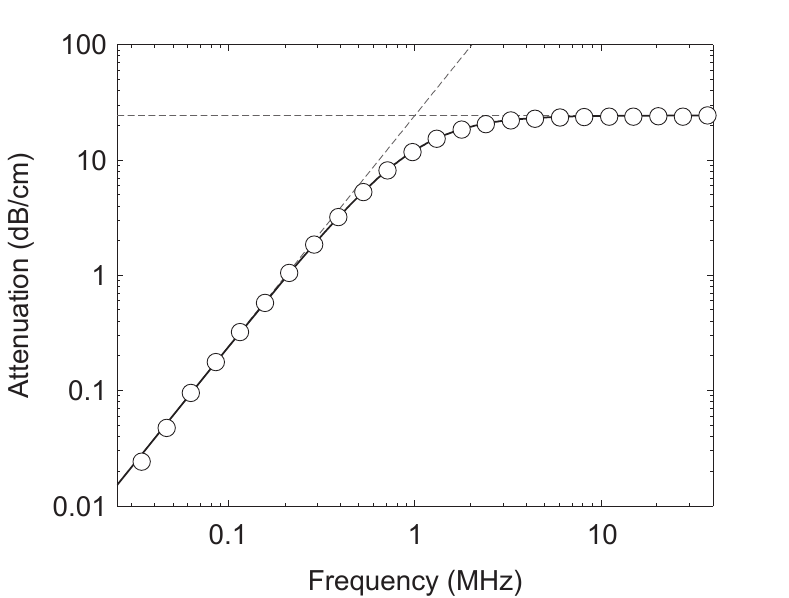}
	\caption{Attenuation retrieved by the numerical algorithm for single relaxation process (circles), theoretical attenuation due to single relaxation (continuous line), for a single relaxing media. The low $\alpha(\omega ) \simeq {\eta _n}{\omega ^2}\tau /2{c_0}$ and high $\alpha = {\eta _n}/2{c_0}{\tau _n}$ frequency limit attenuation are plotted as dash-dotted lines.}
	\label{fig2_attenuation}
\end{figure}

For small amplitude acoustic perturbations in a weakly dispersive media, the phase velocity due to the relaxation processes is related to the attenuation through the Kramers-Kronig relations \cite{ODonnell1981}. Thus, the theoretical phase velocity can be predicted for relaxing media as \cite{Pierce1989}
\begin{equation}\label{eq:c_t}
	{c_p}(\omega ) = {c_0}\left( {1 + \sum\limits_{n = 1}^N {\frac{{{\eta _n}}}{2}\frac{{{\omega ^2}\tau _n^2}}{{1 + {\omega ^2}\tau _n^2}}} } \right).
\end{equation}
On the other hand, in order to study the propagation speed for each spectral component of the numerical method, the phase velocity is computed as
\begin{equation}\label{eq:c_num}
	{c_p}{(\omega)_{fdtd}} = \frac{{w\cdot\left( {{z_1} - {z_0}} \right)}}{{\arg \left( {P(\omega,{z_1})/P(\omega,{z_0})} \right)}}.
\end{equation}
Using the same simulation parameters, Fig. 3 shows the retrieved phase velocity and the theoretical dispersive response of the single relaxation process. The agreement between the model solution and the analytic prediction is again excellent, demonstrating that the inclusion of relaxation processes in nonlinear equations by means of the proposed model exhibit attenuation and dispersion in a correct way. However, typical FDTD numerical dispersion can be observed in the high frequency limit ($f>20$ MHz), where the cumulative phase error lead to phase velocity mismatching. This error can be mitigated in the present numerical algorithm by increasing the number of elements per wavelength.

\begin{figure}[htbp]
	\centering
	\includegraphics[width=8cm]{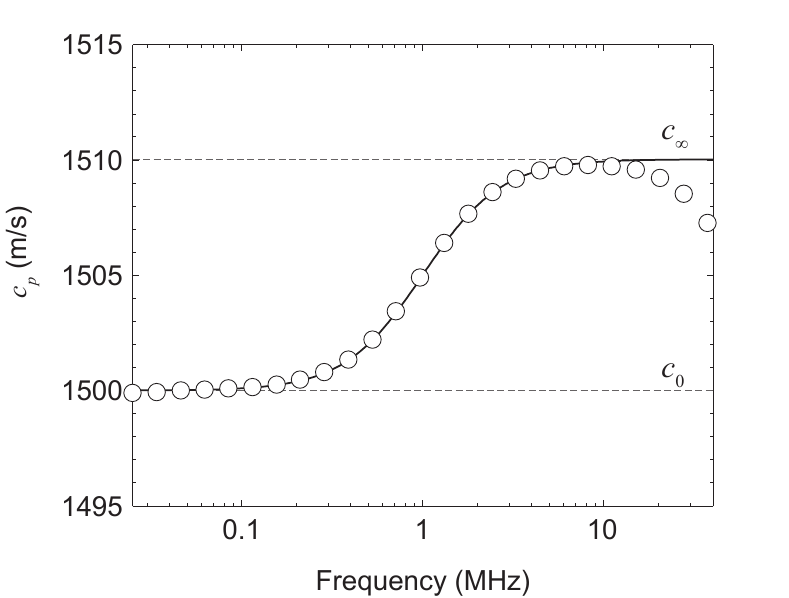}
	\caption{Phase velocity retrieved by the numerical algorithm (circles) and theoretical phase velocity (continuous line) for a single relaxing media.}
	\label{fig3_dispersion}
\end{figure}

\subsection{Frequency power law attenuation}\label{s:validation:power}
An optimization algorithm has been used to fit the numerical attenuation response due to multiple relaxation, Eq. (\ref{eq:alpha_t}), to a frequency power law $\alpha (f)=\alpha _0 f^\gamma$. In order to find the proper relaxation coefficients, this algorithm uses the $fmincon$ function in the optimization toolbox in MATLAB v7.13. Thus, an optimization of the relaxation times $\tau _n$ and relaxation modulus $\eta _n$ values has been done in order to minimize the relative error between the target power law and the computed attenuation. Similar results and accuracy, but longer computational times, where obtained by using a genetic algorithm for the relaxation coefficients optimization. Plane wave propagation was considered and simulation parameters were $c_0=1500$ m/s, $\rho _0=1000$ kg/m$^3$, $B/A=5$, $\eta=8.90\cdot 10^{-4}$ Pa$\cdot$s, $f_0=1$ MHz, $\Delta r = \Delta z = 1.3 \cdot {10^{ - 5}}$ m, $\Delta t = 5.4 \cdot {10^{ - 9}}$ s; that leads to 26 elements per wavelength and a CFL number of 0.9. Only two independent relaxation processes were employed in this section to obtain the target frequency power laws.

Following the above procedure, the relaxation times $\tau _n$ and relaxation modulus $\eta _n$ were optimized for different frequency power laws covering the range of that observed in tissues $\gamma=[1, 1.3, 1.6, 2]$. The attenuation coefficient $\alpha _0$ was chosen for each fit to present an attenuation $\alpha=10$ dB/cm at 10 MHz. The fitting was developed over the typical frequency range for medical ultrasound applications, i. e. 1 to 20 MHz. The results for the attenuation curves are plotted in Fig. 4, where the theoretical and the numerical predictions agree over the frequency range used for the fitting.

\begin{figure}[htbp]
	\centering
	\includegraphics[width=8cm]{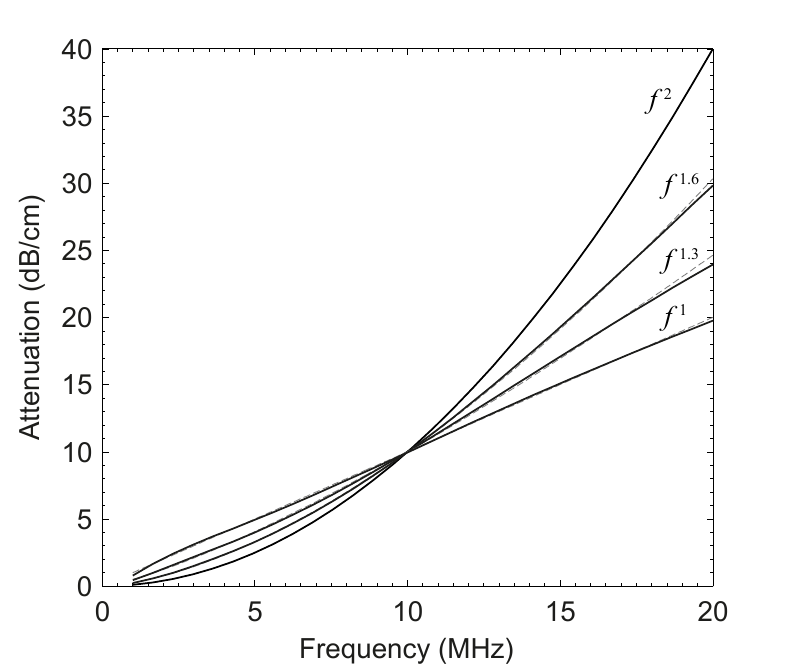}
	\caption{Attenuation retrieved by the numerical algorithm (solid lines) and target frequency power law attenuation (dashed grey lines). By using the optimization algorithm the relaxation times and modulus were optimized for minimize the relative error between the target power laws of $\gamma=[1, 1.3, 1.6, 2]$ and the attenuation retrieved.}
	\label{fig4_curves_solid}
\end{figure}

The fitted exponents for each curve and the goodness of the fit are listed in Table \ref{table:1:fit}, showing that two relaxation process are enough to correctly exhibit a power law in a relative wide frequency range.

\begin{table}[htbp]
	\caption{Results for the fitting procedure for different power frequency attenuation laws}
	\label{table:1:fit}
	\begin{tabular}{l l l l l }
			\hline
				Target $\gamma$ 	& 	Fitted $\gamma$	& 	Target $\alpha _0$  	& 	Fitted $\alpha _0$ 	&	$R^{2}$\\

				 &  & 	 ({Np/m})  	& 	 ({Np/m})	&	 \\
			\hline
				1		& 	1.000	& 	$1.1493\cdot 10^{-5}$ 		& 	$ 1.1521\cdot 10^{-5}$ & 	0.9997\\
	
				1.3		& 	1.279	& 	$1.2800\cdot 10^{-7}$ 		& 	$9.1512\cdot 10^{-8}$ & 	0.9995\\
	
				1.6		&	1.583	& 	$ 9.6526\cdot 10^{-10}$ 		& 	$7.2691\cdot 10^{-10}$ & 	0.9999\\
	
				2		& 	2	& 	$1.1523\cdot 10^{-12}$	& 	$1.1521\cdot 10^{-12}$		& 	0.9999\\
			\hline
		\end{tabular}
\end{table}

\subsection{Fitting attenuation for tissue experimental data}\label{s:validation:exp}
Although a frequency power law dependence can describe the ultrasound attenuation over a finite frequency range, the attenuation data of some particular examples shows variation of the exponent over the entire frequency range \cite{Hill2004}. Thus, as Figure 5 show, the slope of the experimental attenuation data curves for some tissues changes over the measured frequency range. This behavior can be modeled by a sum of relaxation processes by optimizing the relaxation parameters as described above. Thus, the results show that most tissues with locally variable $\gamma$ can be fitted by only a pair of relaxation processes.

\begin{figure}[htbp]
	\centering
	\includegraphics[width=8cm]{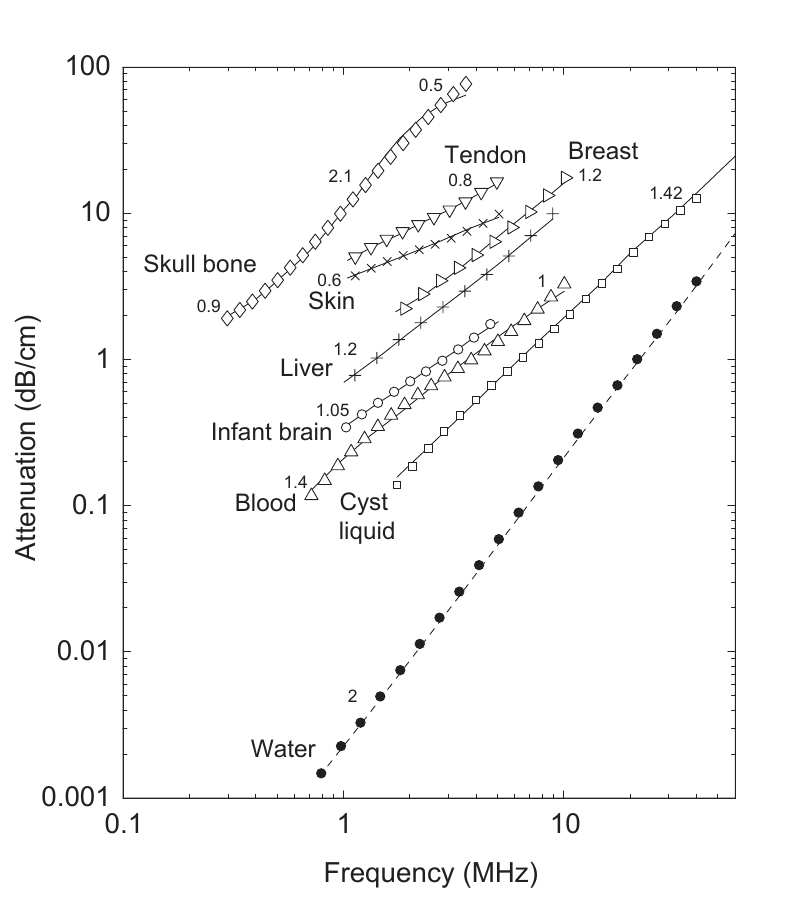}
	\caption{Experimental attenuation data for some tissues \cite{Hill2004} (lines), and obtained by the numerical method (markers) by fitting the parameters of 2 relaxing processes. The numbers above the curves show the coefficient of the frequency power law $\gamma$ for each frequency region (i.e. the slope of the curve).}
	\label{fig5_tissues}
\end{figure}

In this way, Table \ref{table:2:error} shows the error of the numerical attenuation relative to the experimental data. The percent relative error was computed as 
\begin{equation}
	\varepsilon  = \frac{{100}}{{{f_2} - {f_1}}}\int_{{f_1}}^{{f_2}} {\frac{{\left| {{\alpha _e}(f) - \alpha (f)} \right|}}{{{\alpha _e}(f)}}df},
\end{equation}
where $\alpha _e(f)$ is the experimental attenuation data, $f_1$ and $f_2$ define the frequency range of the measurement. As expected, the goodness of fit grows as the number of relaxation processes included increases. However, only two processes are enough to obtain relative errors below 1\% for tissues with $\gamma<2$. In the case of tissues where a local value of $\gamma$ is larger than 2 has been observed the fitting procedure fails, like in the skull bone in the 2 MHz range \cite{Hill2004}. The maximum slope achieved by single relaxation and thermo-viscous losses is $\gamma=2$ for any frequency, so a tissue showing that slope cannot be accurately modeled in this frequency region with the method proposed in this work.

\begin{table}[htbp]
	\caption{Error of the optimized attenuation response relative to the experimental data for $N$ relaxation processes.}
	\begin{tabular}{ l l l l l l}
		\hline
			Tissue 		& 	Power law							&$N=1$				&$N=2$				&$N=3$				&$N=4$		\\
				 		& 										&$\varepsilon(\%)$	&$\varepsilon(\%)$	&$\varepsilon(\%)$	&$\varepsilon(\%)$	\\
		\hline
			Skin	 	& 	$f^{0.6}$							& 6.67				& 0.167				& 0.136				& 0.120		\\
			Liver	 	& 	$f^{1.2}$							& 7.62				& 0.517				& 0.404				& 0.165		\\
			Blood 		& 	$f^{1.4}$, $f^{1}$  				& 8.34				& 0.349				& 0.330				& 0.310		\\
			Breast 		& 	 $f^{0.9}$, $f^{1.2}$				& 5.20 				& 0.216				& 0.209				& 0.205		\\
			Skull bone & 	$f^{0.9}$, $f^{2.1}$, $f^{0.5}$	& 10.60				& 10.54				& 8.628				& 5.189		\\
		\hline
	\end{tabular}
	\label{table:2:error}
\end{table}

Using Kramers-Kronig relations \cite{ODonnell1981}, the variations of sound speed $\Delta c$ can be predicted by the frequency dependent attenuation. Table \ref{table:3:deltac} shows the variation of sound speed observed in the numerical solution over the fitted frequency range. The magnitude of these variations are of the order of magnitude of those measured experimentally in this frequency range, and the frequency dependence observed for the variation is roughly linear as observed in real tissue \cite{Hill2004}. As expected from the relations between dispersion and absorption \cite{ODonnell1981}, the magnitude of the variation in sound speed increases as the total variation of the absorption increases for a frequency range.

\begin{table}[htbp]
	\caption{Variation of sound speed $(\Delta c_0)$ observed numerically for the modeled tissues by means of two relaxation processes and analytical using the Kramers-Kronig relations.}
	\begin{tabular}{ l l l}
		\hline
			Tissue 	& 	Numerical $\Delta c$		& Analytical $\Delta c$\\
				 		& 	$(m/s)$					& $(m/s)$\\
		\hline
			Skull bone & 	80.737						& 70.720\\
			Skin	 	& 	10.148						& 2.460\\
			Breast 	& 	2.323						& 2.455\\
			Liver	 	& 	3.118						& 2.339 \\
			Blood 		& 	0.865						& 0.907 \\
		\hline
	\end{tabular}
	\label{table:3:deltac}
\end{table}

\subsection{Nonlinear regime}\label{s:validation:nonlinear}
In order to study the convergence of the numerical calculations to an analytical solution of the model in the nonlinear regime, a medium with frequency squared dependence attenuation is implemented using the adequate relaxation times and relaxation modulus as explained above. The solution for the frequency squared absorption is compared with the analytical solution for a plane wave traveling through a thermo-viscous fluid proposed by Mendousse \cite{Pierce1989}:
\begin{equation}\label{eq:medousse}
	\frac{p}{{{p_0}}} = \frac{{\frac{4}{\Gamma }\sum\nolimits_{n = 1}^\infty  {{{\left( { - 1} \right)}^{n + 1}}{I_n}\left( {\frac{\Gamma }{2}} \right)n{e^{ - {n^2}\sigma /\Gamma }}\sin \left( {n\omega t'} \right)} }}{{{I_0}\left( {\frac{\Gamma }{2}} \right) + 2\sum\nolimits_{n = 1}^\infty  {{{\left( { - 1} \right)}^n}{I_n}\left( {\frac{\Gamma }{2}} \right)n{e^{ - {n^2}\sigma /\Gamma }}\cos \left( {n\omega t'} \right)} }}
\end{equation}
where $\Gamma$ is the Goldberg number as $\Gamma = {c_0^3}/{b{\omega ^2}\bar x}$. Here the diffusion coefficient $b = {\left(4/3 \eta + \zeta \right)}/\rho _0 $, $\bar x = 1/{\left(\left(1+B/2A\right)M_a k_0\right)}$ is the shock formation distance, the acoustic Mach number $M_a = v_0/c_0$, $v_0$ is the particle velocity excitation at $x = 0$ and the wavenumber $k_0 = 2\pi/\lambda _0$.

\begin{figure}[htbp]
	\centering
	\includegraphics[width=8cm]{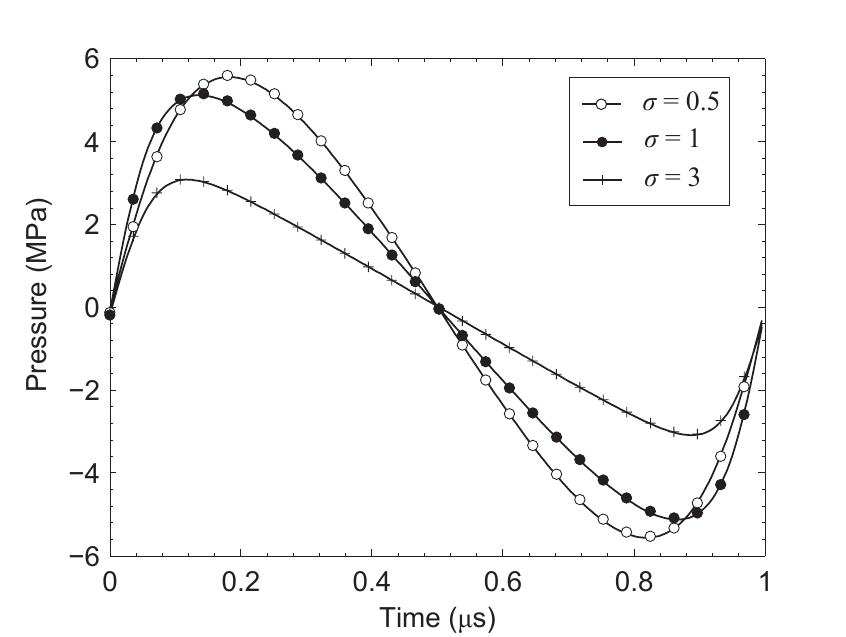}
	\caption{Analytical solution (continuous lines) for plane wave traveling through a viscous fluid (frequency squared power law) and simulated waveforms at $\sigma=0.5$ (white circles), $\sigma=1$ (black circles) and $\sigma=3$ (asterisk).}
	\label{fig6_nonlinear_mendousse}
\end{figure}

In this case, a $f_0=1$ MHz sinusoidal plane wave propagation was considered with an amplitude of 6 MPa, leading to a shock propagation distance of $\bar x=17\lambda$. Thus, defining the distance normalized by the shock formation distance $\sigma  = x/\bar x$, the pressure waveforms $p(t,\sigma)$ for the numerical and the analytic solutions are shown in Figure 6 for $\sigma=(0.5, 1, 3)$. The numerical solution agrees with the predicted by the analytical solution (Eq. \ref{eq:medousse}) at different distances. 

In order to study the accuracy of the algorithm, the amplitude of the first three harmonics has been extracted for the numerical and analytic solutions and plotted versus $\sigma$ in Figure 7. This result shows that the wave steepening due to the nonlinear processes are well resolved by the numerical method presented here. However, as a consequence of the numerical dispersion of the FDTD algorithm, the numerical sound speed does not exactly match the theoretical sound speed, so a phase error is always present on the solution. The observed relative error for the first three harmonic amplitude decreases due to grid coarsening by a square law (i. e. the numerical method is second order accuracy). Thus, the magnitude of the error mainly depends on the traveled propagated distance and the number of elements per wavelength. In the case of a traveled distance of 85 $\lambda$, a grid of 32 elements per wavelength is needed to obtain a relative error below 1\% for the third harmonic. In this case the relative error of the lower harmonics is always lower, where the fundamental harmonic error was 0.072 \%. The grid can be reduced if shorter path length are considered or if tolerance requirements for the higher spectral components are less restrictive, saving computational resources in higher dimensional reference systems (i.e. 3D. Cartesian or cylindrical coordinate systems).

\begin{figure}[htbp]
	\centering
	\includegraphics[width=8cm]{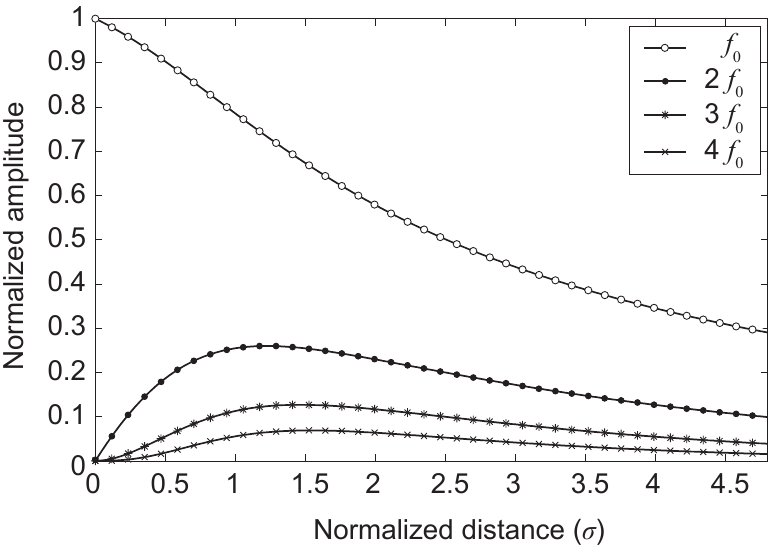}
	\caption{Harmonic amplitude versus normalized distance $\sigma$ for the three first harmonic components of the analytical (continuous lines) and numerical solutions, where the evolution of the fundamental component $f_0$ (white circles), $2f_0$ (black circles), $3f_0$ (asterisks) and $4f_0$ (crosses) agrees with the analytical solution.}
	\label{fig7_harmonics_sigma}
\end{figure}

\section{Conclusions}\label{s:conclusions}
In the present work a model for nonlinear acoustic waves in relaxing media is presented in time-domain formulation which does not require convolutional operators. A numerical solution by means of finite differences in time domain have been obtained, showing that the theoretical attenuation and dispersion due to relaxation processes can be achieved by the numerical method with accuracy. These result can be used to model typical relaxation process (e. g. the processes observed in air, associated to the molecules of oxygen and nitrogen, or in seawater, associated to the relaxation of boric acid and magnesium sulfate). 

Moreover, in this work a method for model a frequency power law on attenuation by means of multiple relaxation was described. After an optimization of the relaxation coefficients, numerical results show that a pair of relaxation processes are enough to obtain frequency power law with accuracy (relative error below one percent) for medical ultrasound predictions. Therefore, only two auxiliary fields are necessary to model a frequency power law attenuation over a typical medical ultrasound frequency band. 

Besides, the proposed method can achieve local variations of the power in the frequency power law, so an arbitrary attenuation curve can be modeled by means of the proper optimization of the relaxation coefficients. The only observed limitation in that the local exponent of the frequency power law must be lower than 2. This feature of the presented method is an advantage when compared with most fractional derivatives methods, where the attenuation follows an exact but unique frequency power law over the entire frequency range. On the other hand, due to the numerical behavior of the finite differences in time domain method, an increase on the dispersion has been observed, mismatching the phase velocity in the higher frequency limit. This effect leads to cumulative phase errors on large propagation distances, so a convenient grid refinement must be employed, increasing the computational cost of the algorithm. Moreover, if path length is on the order of thousand wavelengths and phase error is crucial for a specific prediction, correct dispersion can be accurately achieved by \textit{k}-space or pseudospectral numerical methods saving computational time and memory.

\begin{acknowledgments}\label{acknowledgments}
The work was supported by Spanish Ministry of Science and Innovation through project FIS2011-29731-C02-01. N. Jiménez acknowledges financial support from the Universitat Politècnica de València (Spain) through the FPI-2011 PhD grant.
\end{acknowledgments}

\end{document}